\documentstyle[twoside,12pt,epsf]{article}

\catcode`\@=11
\long\def\@makefntext#1{
\protect\noindent \hbox to 3.2pt {\hskip-.9pt  
$^{{\eightrm\@thefnmark}}$\hfil}#1\hfill}		%CAN BE USED 

\def\thefootnote{\fnsymbol{footnote}}
\def\@makefnmark{\hbox to 0pt{$^{\@thefnmark}$\hss}}	%ORIGINAL 
	
\def\ps@myheadings{\let\@mkboth\@gobbletwo
\def\@oddhead{\hbox{}
\rightmark\hfil\eightrm\thepage}   
\def\@oddfoot{}\def\@evenhead{\eightrm\thepage\hfil
\leftmark\hbox{}}\def\@evenfoot{}
\def\sectionmark##1{}\def\subsectionmark##1{}}

\oddsidemargin=\evensidemargin
\renewcommand{\thefootnote}{\fnsymbol{footnote}}

\newcounter{sectionc}\newcounter{subsectionc}\newcounter{subsubsectionc}
\renewcommand{\section}[1] {\vspace{12pt}\addtocounter{sectionc}{1} 
\setcounter{subsectionc}{0}\setcounter{subsubsectionc}{0}\noindent 
	{\tenbf\thesectionc. #1}\par\vspace{5pt}}
\renewcommand{\subsection}[1] {\vspace{12pt}\addtocounter{subsectionc}{1} 
	\setcounter{subsubsectionc}{0}\noindent 
	{\bf\thesectionc.\thesubsectionc. {\kern1pt \bfit #1}}\par\vspace{5pt}}
\renewcommand{\subsubsection}[1] {\vspace{12pt}\addtocounter{subsubsectionc}{1}
	\noindent{\tenrm\thesectionc.\thesubsectionc.\thesubsubsectionc.
	{\kern1pt \tenit #1}}\par\vspace{5pt}}
\newcommand{\nonumsection}[1] {\vspace{12pt}\noindent{\tenbf #1}
	\par\vspace{5pt}}

\newcounter{appendixc}
\newcounter{subappendixc}[appendixc]
\newcounter{subsubappendixc}[subappendixc]
\renewcommand{\thesubappendixc}{\Alph{appendixc}.\arabic{subappendixc}}
\renewcommand{\thesubsubappendixc}
	{\Alph{appendixc}.\arabic{subappendixc}.\arabic{subsubappendixc}}

\renewcommand{\appendix}[1] {\vspace{12pt}
        \refstepcounter{appendixc}
        \setcounter{figure}{0}
        \setcounter{table}{0}
        \setcounter{lemma}{0}
        \setcounter{theorem}{0}
        \setcounter{corollary}{0}
        \setcounter{definition}{0}
        \setcounter{equation}{0}
        \renewcommand{\thefigure}{\Alph{appendixc}.\arabic{figure}}
        \renewcommand{\thetable}{\Alph{appendixc}.\arabic{table}}
        \renewcommand{\theappendixc}{\Alph{appendixc}}
        \renewcommand{\thelemma}{\Alph{appendixc}.\arabic{lemma}}
        \renewcommand{\thetheorem}{\Alph{appendixc}.\arabic{theorem}}
        \renewcommand{\thedefinition}{\Alph{appendixc}.\arabic{definition}}
        \renewcommand{\thecorollary}{\Alph{appendixc}.\arabic{corollary}}
        \renewcommand{\theequation}{\Alph{appendixc}.\arabic{equation}}
        \noindent{\tenbf Appendix \theappendixc #1}\par\vspace{5pt}}
\newcommand{\subappendix}[1] {\vspace{12pt}
        \refstepcounter{subappendixc}
        \noindent{\bf Appendix \thesubappendixc. {\kern1pt \bfit #1}}
	\par\vspace{5pt}}
\newcommand{\subsubappendix}[1] {\vspace{12pt}
        \refstepcounter{subsubappendixc}
        \noindent{\rm Appendix \thesubsubappendixc. {\kern1pt \tenit #1}}
	\par\vspace{5pt}}

\newcommand{\textlineskip}{\baselineskip=15pt}
\newcommand{\smalllineskip}{\baselineskip=10pt}

\def\eightcirc{
\begin{picture}(0,0)
\put(4.4,1.8){\circle{6.5}}
\end{picture}}
\def\eightcopyright{\eightcirc\kern2.7pt\hbox{\eightrm c}}

\def\abstracts#1#2#3{{
	\centering{\begin{minipage}{4.5in}\baselineskip=10pt\footnotesize
	\parindent=0pt #1\par 
	\parindent=15pt #2\par
	\parindent=15pt #3
	\end{minipage}}\par}} 

\renewenvironment{thebibliography}[1]
	{\frenchspacing
	 \ninerm\baselineskip=11pt
	 \begin{list}{\arabic{enumi}.}
	{\usecounter{enumi}\setlength{\parsep}{0pt}
	 \setlength{\leftmargin 12.7pt}{\rightmargin 0pt} %FOR 1--9 ITEMS
	 \setlength{\itemsep}{0pt} \settowidth
	{\labelwidth}{#1.}\sloppy}}{\end{list}}

\newcounter{itemlistc}
\newcounter{romanlistc}
\newcounter{alphlistc}
\newcounter{arabiclistc}

\newcommand{\fcaption}[1]{
        \refstepcounter{figure}
        \setbox\@tempboxa = \hbox{\footnotesize Fig.~\thefigure. #1}
        \ifdim \wd\@tempboxa > 5in
           {\begin{center}
        \parbox{5in}{\footnotesize\smalllineskip Fig.~\thefigure. #1}
            \end{center}}
        \else
             {\begin{center}
             {\footnotesize Fig.~\thefigure. #1}
              \end{center}}
        \fi}

\newcommand{\tcaption}[1]{
        \refstepcounter{table}
        \setbox\@tempboxa = \hbox{\footnotesize Table~\thetable. #1}
        \ifdim \wd\@tempboxa > 5in
           {\begin{center}
        \parbox{5in}{\footnotesize\smalllineskip Table~\thetable. #1}
            \end{center}}
        \else
             {\begin{center}
             {\footnotesize Table~\thetable. #1}
              \end{center}}
        \fi}

\def\@citex[#1]#2{\if@filesw\immediate\write\@auxout
	{\string\citation{#2}}\fi
\def\@citea{}\@cite{\@for\@citeb:=#2\do
	{\@citea\def\@citea{,}\@ifundefined
	{b@\@citeb}{{\bf ?}\@warning
	{Citation `\@citeb' on page \thepage \space undefined}}
	{\csname b@\@citeb\endcsname}}}{#1}}

\newif\if@cghi
\def\cite{\@cghitrue\@ifnextchar [{\@tempswatrue
	\@citex}{\@tempswafalse\@citex[]}}
\def\citelow{\@cghifalse\@ifnextchar [{\@tempswatrue
	\@citex}{\@tempswafalse\@citex[]}}
\def\@cite#1#2{{$\null^{#1}$\if@tempswa\typeout
	{IJCGA warning: optional citation argument 
	ignored: `#2'} \fi}}

\def\pmb#1{\setbox0=\hbox{#1}
	\kern-.025em\copy0\kern-\wd0
	\kern.05em\copy0\kern-\wd0
	\kern-.025em\raise.0433em\box0}

\def\fnm#1{$^{\mbox{\scriptsize #1}}$}
\def\fnt#1#2{\footnotetext{\kern-.3em
	{$^{\mbox{\scriptsize #1}}$}{#2}}}

\def\fpage#1{\begingroup
\voffset=.3in
\thispagestyle{empty}\begin{table}[b]\centerline{\footnotesize #1}
	\end{table}\endgroup}

\def\runninghead#1#2{\pagestyle{myheadings}
\markboth{{\protect\footnotesize\it{\quad #1}}\hfill}
{\hfill{\protect\footnotesize\it{#2\quad}}}}
\font\tenrm=cmr10
\font\tenit=cmti10 
\font\tenbf=cmbx10
\font\bfit=cmbxti10 at 10pt
\font\ninerm=cmr9

\font\eightrm=cmr8

\def\qed{\hbox{${\vcenter{\vbox{			%HOLLOW SQUARE
   \hrule height 0.4pt\hbox{\vrule width 0.4pt height 6pt
   \kern5pt\vrule width 0.4pt}\hrule height 0.4pt}}}$}}

\renewcommand{\thefootnote}{\fnsymbol{footnote}}	%USE SYMBOLIC FOOTNOTE

\begin{document}

\runninghead{D.J. Fern\'andez C.} {SUSUSY quantum mechanics} 

\normalsize\textlineskip
\thispagestyle{empty}
\setcounter{page}{1}

%\copyrightheading{}			%{Vol. 0, No. 0 (1993) 000--000}

\vspace*{0.0truein}

\fpage{1}
\centerline{\bf SUSUSY QUANTUM MECHANICS}
\vspace*{0.37truein}
\centerline{\footnotesize DAVID J. FERN\'ANDEZ C.}
\vspace*{0.015truein}
\centerline{\footnotesize\it Depto. de F\'{\i}sica, CINVESTAV}
\baselineskip=10pt
\centerline{\footnotesize\it A.P. 14-740, 07000 M\'exico D.F., M\'exico}
\vspace*{0.225truein}
%\publisher{(received date)}{(revised date)}

\vspace*{0.21truein}
\abstracts{The exactly solvable eigenproblems in Schr\"odinger quantum 
mechanics typically involve the differential ``shift operators". In the 
standard supersymmetric (SUSY) case, the shift operator turns out to be of 
first order. In this work, I discuss a technique to generate exactly solvable 
eigenproblems by using second order shift operators. The links between this 
method and SUSY are analysed. As an example, we show the existence of a 
two-parametric family of exactly solvable Hamiltonians, which contains the 
Abraham-Moses potentials as a particular case. }{}{} 

%\textlineskip			%) USE THIS MEASUREMENT WHEN THERE IS
%\vspace*{12pt}			%) NO SECTION HEADING

\vspace*{1pt}\textlineskip	%) USE THIS MEASUREMENT WHEN THERE IS
\section{\bf Introduction}	%) A SECTION HEADING
\vspace*{-0.5pt}
\noindent
The number of exactly solvable eigenproblems in non-relativistic quantum 
mechanics is small, and most of them can be dealt with the factorization 
method. This technique, introduced long ago by Schr\"odinger,${}^1$ was 
analysed in depth by Infeld and Hull,${}^2$ who made an exhaustive 
classification of factorizable potentials. Later on, Witten noticed the 
possibility of arranging the Schr\"odinger's Hamiltonians into isospectral 
pairs (supersymmetric (SUSY) partners).${}^3$ The resulting {\it 
supersymmetric quantum mechanics} catalysed the study of hierarchies of 
`exactly solvable Hamiltonians'. An additional step was Mielnik's `atypical' 
factorization${}^4$ through which the general SUSY partner for the oscillator 
was found; this technique was immediately applied to the hydrogen 
potential.${}^5$ Meanwhile, Nieto${}^6$ and Andrianov {\it et. al.}${}^7$ 
put the method on its natural background discovering the links between SUSY, 
factorization and Darboux algorithm. These developments caused the renaissance 
of factorization and related algebraic methods, with particular attention 
focused on the first order differential shift operators.${}^{8-16}$ 

As can be noticed, however, the scheme is still narrow. An obvious 
generalization arises when higher order differential shift operators are used 
to connect the Hamiltonian pair. The idea of HSUSY (higher order SUSY), 
recently put forward by Andrianov {\it et. al.}${}^{17-18}$ (see also 
${}^{19}$), incubated since 70-tieth.${}^{20-21}$ In this paper we will 
restrict ourselves to the case when the shift operator is of second order, and 
we name it SUSUSY. 

\pagebreak

\textheight=7.8truein
\setcounter{footnote}{0}
\renewcommand{\thefootnote}{\alph{footnote}}

\section{\bf Second order shift operator technique}
\noindent
We postulate the existence of a second order differential operator 
interconnecting two different Hamiltonians $H, \ {\widetilde H}$: 
\begin{equation}
{\widetilde H} A^\dagger = A^\dagger H, \label{(1)}
\end{equation}
\begin{equation}
H = - {d^2\over dx^2} + V(x), \quad {\widetilde H} = - {d^2\over dx^2} + 
{\widetilde V}(x), \label{(2)}
\end{equation}
\begin{equation}
A^\dagger = {d^2\over dx^2} + \beta(x) {d \over dx} + \gamma(x).
\label{(3)}
\end{equation}
Equality (1) imposes some restrictions to the functions $\{V(x),{\widetilde 
V}(x),\beta(x),\gamma(x)\}$: 
\begin{equation}
{\widetilde V} = V + 2\beta', \quad
2V+\delta = \beta^2 - 2\gamma -\beta' ,  \quad
V'' + \beta V' = 2 \gamma \beta' -\gamma'',  \label{(4)}
\end{equation}
where $\delta$ is an integration constant. We shall suppose that $\beta(x)$ is 
given and we shall express the other functions $\{ V(x), \ {\widetilde V}(x), 
\ \gamma(x)\}$ in terms of it. If we solve $\gamma(x)$ from second equation 
(4) and substitute the result in the third equation (4), we get: 
\begin{equation}
{\beta''' \over 2} -2 \beta'^2 + \beta'\beta^2 - \beta \beta'' -\delta\beta' =
\beta V' + 2 \beta' V . \label{(5)}
\end{equation}
Multiplying by $\beta$, it can be immediately integrated, yielding:
\begin{equation}
V(x)= {\beta''\over 2\beta}-\left({\beta'\over 2\beta}\right)^2 - \beta' +
{\beta^2\over 4} + {c\over \beta^2} - {\delta\over 2}, \label{(6)}
\end{equation}
where $c$ is a new integration constant. The other two unknown functions 
become:
\begin{equation}
{\widetilde V}(x)= {\beta''\over 2\beta}-\left({\beta'\over 2\beta}\right)^2 + 
\beta' + {\beta^2\over 4} + {c\over \beta^2} - {\delta\over 2}, \quad
\gamma(x) = -{\beta''\over 2\beta}+\left({\beta'\over 2\beta}\right)^2 + 
{\beta' \over 2} + {\beta^2\over 4} - {c\over \beta^2} . \label{(7)} 
\end{equation} 
Before going to the particular cases, let us notice a curious relation between 
the second order shift operator and Witten idea of the SUSY quantum mechanics. 

\section{\bf Second order SUSY (SUSUSY)}
\noindent
According to Witten,${}^3$ SUSY arises by defining a set of operators $Q_i$ 
that commute with the (supersymmetric) Hamiltonian $H_s$, 
\begin{equation}
[Q_i,H_s]=0, \quad i=1\cdots N,  \label{(8)} 
\end{equation}
and satisfy the algebra
\begin{equation}
\{Q_i,Q_j\} = \delta_{ij} H_s,  \label{(9)}
\end{equation}
where $[\cdot,\cdot]$ represents the commutator and $\{\cdot,\cdot\}$ the 
anticommutator. Now, with the aid of the operators $A^\dagger, \ A$ of the 
previous section, one can construct a case of the supersymmetric algebra 
(8-9) with $N=2$. 

With this aim, define the supercharges:${}^{17-18}$
\begin{equation}
Q=\left(\matrix{0 & 0 \cr A & 0} \right), \qquad Q^\dagger = \left(\matrix{0 
& A^\dagger \cr 0 & 0} \right), \label{(10)}
\end{equation}
where $A^\dagger$ is given in (3) and $A$ is its adjoint. Notice that $Q^2 
= Q^{\dagger 2} = 0$. Let us construct an operator, which we cannot abstain to 
call the SUSUSY `Hamiltonian': 
\begin{equation}
H_{ss} = \{ Q, Q^\dagger \} = \left( \matrix{ A^\dagger A & 0 \cr 0 & 
AA^\dagger} \right) = \left( \matrix{H^+ & 0 \cr 0 & H^-}\right). \label{(11)}
\end{equation}
Using the SUSY languaje, $H^+=A^\dagger A$ and $H^-=AA^\dagger$ should be 
called the SUSY partners. Notice that the SUSUSY `Hamiltonian' $H_{ss}$ 
commutes with the two supercharges $Q$ and $Q^\dagger$. The SUSY generators 
$Q_1= (Q^\dagger + Q)/\sqrt{2}$, $Q_2=(Q^\dagger - Q)/i\sqrt{2}$, and $H_{ss}$ 
satisfy the supersymmetric algebra (8-9). 

Notice that the SUSY partners $H^+,H^-$ are now the fourth order differential 
operators. It can be shown that $H^+$ commutes with ${\widetilde H}$ and $H^-$ 
commutes with $H$. Hence, $H^+$ can be a certain function of ${\widetilde H}$ 
and $H^-$  a function of $H$. Indeed: 
\begin{equation}
H^+ = \left({\widetilde H} + {\delta \over 2}\right)^2 -c, \quad 
H^- = \left(H + {\delta \over 2}\right)^2 -c. \label{(12)}
\end{equation}
A physical Hamiltonian $H_s$ can be defined (compare with the recent ideas of 
${}^{17-18}$), 
\begin{equation}
H_s = \left( \matrix{ {\widetilde H} & 0 \cr 0 & H}\right), \label{(13)}
\end{equation}
and the SUSUSY `Hamiltonian' $H_{ss}$ is related to $H_s$ by means of: 
\begin{equation}
H_{ss} = \left( H_s + {\delta \over 2}\right)^2 - c. \label{(14)}
\end{equation}
Thus, the SUSUSY `Hamiltonian' $H_{ss}$ is a quadratic form of a physical 
Hamiltonian $H_s$. The diagonal elements of $H_s$ are the two Hamiltonians $H 
, \ {\widetilde H}$ of the previous section, which are related by the second 
order differential operators $A, \ A^\dagger$ (compare ${}^{17-18}$). 

\newpage
\section{\bf Example: the SUSUSY oscillator}
\noindent
We shall now look for the SUSUSY analogue of the oscillator Hamiltonian 
\begin{equation}
H = -{d^2\over dx^2} + x^2  \label{(15)}
\end{equation}
We will try to show the existence of a 2-parametric family of potentials 
isospectral to $V(x) = x^2$. This has to do with the general solution 
$\beta(x)$ of equation (6). This solution of course should include the  ladder 
operator $A^\dagger = (a^\dagger)^2$, where $a^\dagger = - d/dx +x$ is the 
standard ladder operator of the harmonic oscillator. This means that for $V(x) 
= x^2$ and $\beta_p(x) = -2 x$, equation (6) should become an identity, which 
fixes the constants to $c=1, \delta=4$. Substituting these results again in 
(6) and multiplying by $2\beta^2$, we get: 
\begin{equation} 
\beta\beta'' - {\beta'^2\over 2} - 2\beta^2\beta' + {\beta^4\over 2} - 
4\beta^2 -2 x^2 \beta^2 + 2 = 0.  \label{(16)}
\end{equation} 
Let us notice the existence of an explicit solution more general than $\beta_p 
= -2 x$. It arises after multiplying the standard raising operator $a^\dagger$ 
by the operator $b^\dagger$ of atypical factorizations,${}^4$ i.e., $A^\dagger 
= b^\dagger a^\dagger$, leading to: 
\begin{equation} 
\beta_p(x) = -2x - {e^{-x^2}\over \lambda + \int_0^x e^{-y^2} dy}, \quad
\vert\lambda\vert > {\sqrt{\pi}\over 2}. \label{(17)} 
\end{equation} 
The general solution of (16), which depends on two constants, should reduce 
itself to (17) as one of them takes a particular value (or one of them becomes 
a function of the other one). 

Here, I would like to present some partial numeric results obtained when (16) 
is integrated to provide $\beta(x)$ which arises as continuous deformations of 
the particular solutions (17). With this aim we choose the initial point 
$(\beta(0),\beta'(0))$ on the Poincar\'e plane close to 
$(\beta_p(0),\beta_p'(0))$ and use then a standard numeric integration 
package\fnm{a}\fnt{a}{We have employed the routine `NDSolve' of `Mathematica'} 
to find $\beta(x)$ and to look for singularities in the corresponding SUSUSY 
potential ${\widetilde V}(x) = x^2 + 2\beta'(x)$. If there is no singularity, 
we increase slightly $\beta'(0)$ maintaining $\beta(0)$ fixed, and repeat the 
integration until finding the upper threshold of $\beta'(0)$: above this 
threshold a singularity arises while below it disappears. A similar procedure 
is used to find the low threshold. After that, we make a small change of 
$\beta(0)$ along $(\beta_p(0),\beta_p'(0))$ and start again the whole  
process. In this way we can split the $\beta\beta'$-plane into the region 
where the SUSUSY potential ${\widetilde V}(x)$ is free of singularities and 
the rest. Notice that the points $(\beta_p(0),\beta_p'(0))$ provide a curve on 
$\beta\beta'$-plane: 
\begin{equation}
\beta_p'(0) = -2 + \beta_p^2(0), \quad \vert\beta_p(0)\vert < 
{2\over\sqrt\pi}. \label{(18)} 
\end{equation}
Departing from (18) we made the classification on figure 1. 

\begin{figure}[htbp]
\vspace*{13pt}
\begin{minipage}{10truecm}
\hspace*{3truecm}
\epsfxsize=10truecm
\epsfbox{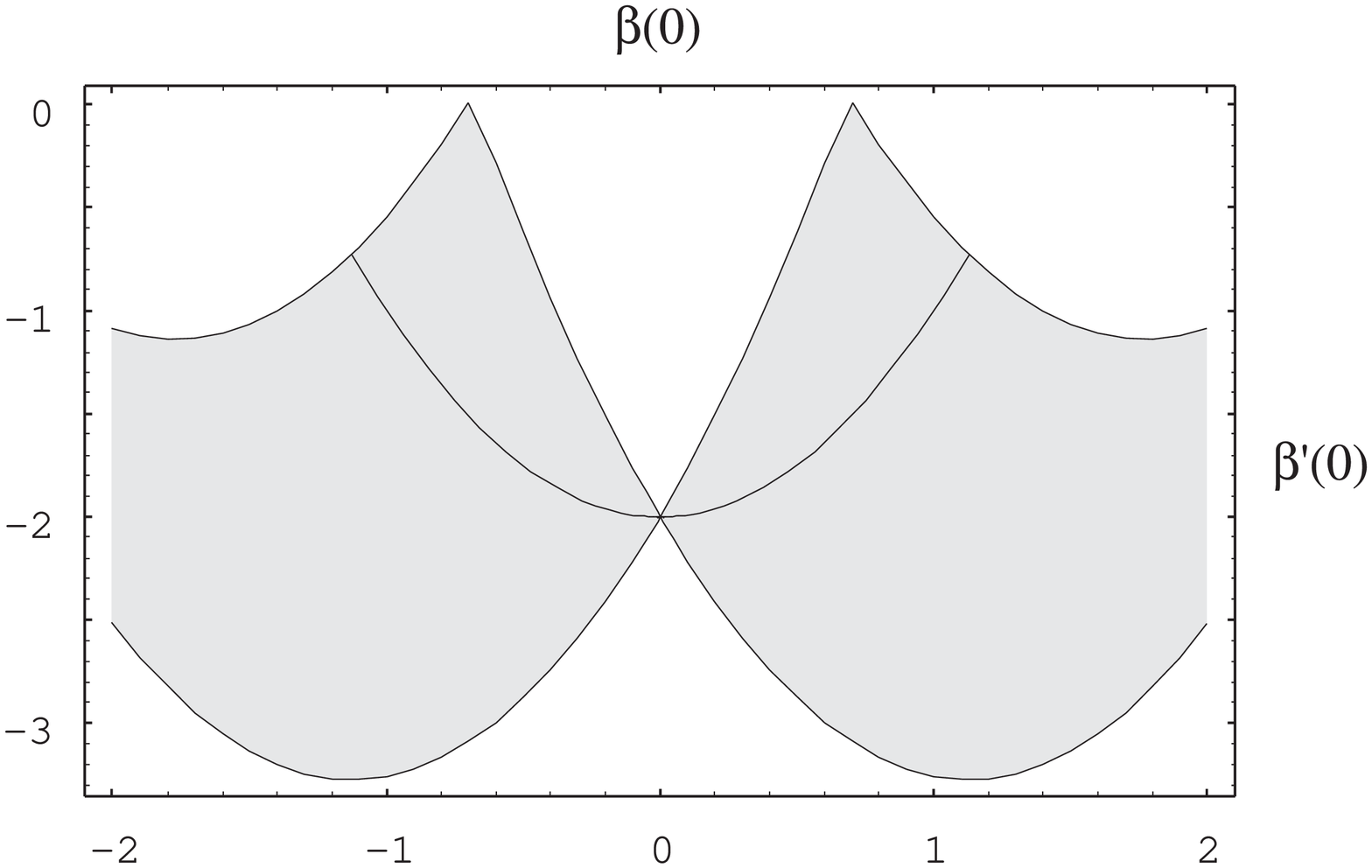}
\end{minipage}
\vspace*{13pt}
\fcaption{Classification of the $\beta\beta'$-plane into regions where the 
SUSUSY potential ${\widetilde V}(x)$ has no singularities (the shadowed 
regions) and those with singularity (the white regions).} 
\end{figure}

As we can see, there is a non-trivial region, the shadowed one, where the 
SUSUSY potential ${\widetilde V}(x)$ has no singularity; it comprises the 
curve in (18). For the purposes of this paper, to show that the general 
family of SUSUSY oscillator potentials is two parametric, our calculation 
brings already some information. 

The SUSUSY potentials ${\widetilde V}(x)$ for various values of the pair 
$(\beta(0), \beta'(0))$ lying at the shadowed region (no singularity, we have 
fixed $\beta(0) = -0.7$) are shown in figure 2\fnm{b}\fnt{b}{Indeed, 
${\widetilde V}(x)$ is displaced with respect to $V(x)=x^2$ a quantity $\delta 
E = -4$. This can be seen after realizing that ${\widetilde V}(x) + 4 
\rightarrow x^2$ when $(\beta(0),\beta'(0))\rightarrow (0,-2)$. Hence, we 
decided to represent on the vertical axis of figure 2 the potentials 
${\widetilde V}(x)+4$ rather that ${\widetilde V}(x)$.}. This family is richer 
than the Abraham-Moses (AM) SUSY potentials:${}^{4,22}$ 
\begin{equation}
V_\lambda(x) = x^2 - 2 {d\over dx}\left({e^{-x^2}\over \lambda + \int_0^x e^{-
y^2}dy}\right). \label{(19)}
\end{equation}
This is so because ${\widetilde V}(x)$ is essentially two-parametric while 
(19) is just one-parame\-tric. Indeed, the $V_\lambda(x)$ of (19) can be 
numerically reconstructed by solving (16) with the points of (18) as the 
initial conditions. The SUSUSY family obtained by this procedure coincides 
with a plot of the analytic results (19). 

\begin{figure}[htbp]
\vspace*{13pt}
\begin{minipage}{10truecm}
\hspace*{3truecm}
\epsfxsize=10truecm
\epsfbox{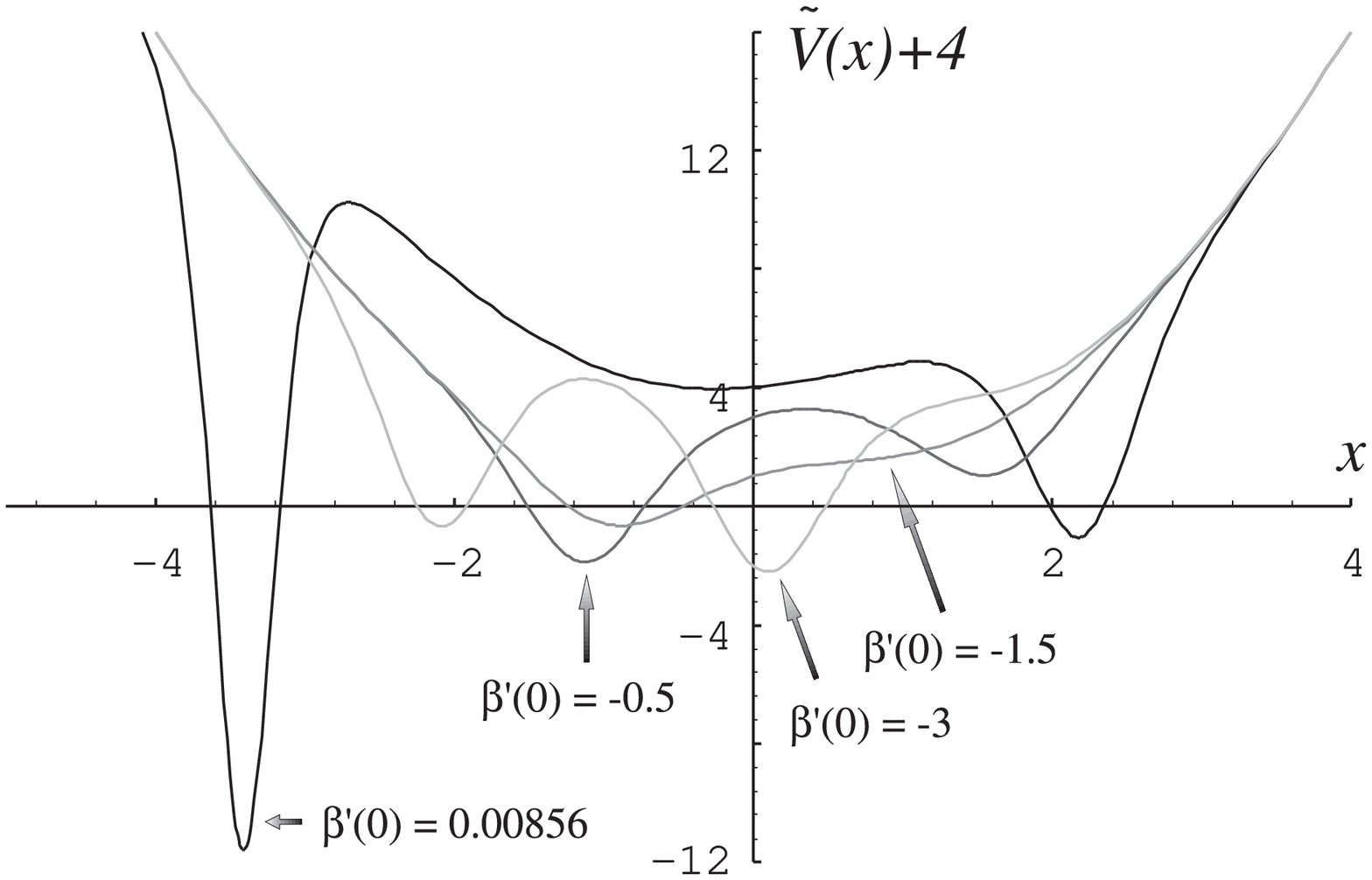}
\end{minipage}
\vspace*{13pt}
\fcaption{The SUSUSY potentials ${\widetilde V}(x)+4$ for some values of 
$\beta'(0)$ and $\beta(0) = -0.7$. All pairs $(\beta(0), \beta'(0))$ lie in 
the region where there is no singularity for ${\widetilde V}(x)$.} 
\end{figure}

Interesting that the SUSUSY family ${\widetilde V}(x)$ embraces some cases of 
the widely discussed {\it double well potentials} (DWP). The dynamics of a 
system in these potentials is of some relevance, because it illustrates the 
differences between the classical and quantum regimes. In particular, it 
clearly shows one of the most intriguing quantum effects, the tunneling of the 
system from one well to the other as a result of the evolution. In most of the 
situations where a DWP is a SUSY pair of the oscillator potential, the DWP 
spectrum has one level more below the ground state energy of the oscillator. 
Moreover, it is usually symmetric with respect to $x=0$ (see e.g. 
${}^{9,23}$). For our SUSUSY DWP, apparently, it is unneccessary to add any 
level below the ground state energy of the oscillator to generate the double 
well: the spectra of ${\widetilde V}(x) + 4$ and $V(x) = x^2$ are 
equal.\fnm{c}\fnt{c}{This is at the moment a hypothesis supported by our 
numerical plots of $\beta(x)$ and ${\widetilde V}(x)$. (The continuity 
argument might be important.)} The price to pay is that ${\widetilde V}(x)$ 
and $V(x) = x^2$ are not precisely the SUSY partners, as shown in section 3. 
Moreover, although ${\widetilde V}(x)$ is a double well, it turns out that it 
is not symmetric with respect to any point $x=x_0$. We hope that the SUSUSY 
treatment here presented can be implemented to other physically interesting 
potentials. 

\nonumsection{\bf Acknowledgements}
\noindent
The support of CONACYT is acknowledged. 

\newpage

\nonumsection

\end{document}